\documentclass[twocolumn,showpacs,preprintnumbers,superscriptaddress]{revtex4}
\usepackage{amsmath}
\usepackage{amsfonts}
\usepackage{dcolumn}
\usepackage{bm}
\usepackage{epsfig}
\usepackage{epstopdf}
\usepackage{graphicx}
\usepackage{amsfonts}
\usepackage{amssymb}
\usepackage{mathrsfs}
\usepackage{color}
\usepackage{appendix}
\usepackage{multirow}
\usepackage{threeparttable}

\begin{document}

\title{An efficient cooling of the quantized vibration by a four-level
configuration}
\author{Lei-Lei Yan}
\affiliation{State Key Laboratory of Magnetic Resonance and Atomic and Molecular Physics,
Wuhan Institute of Physics and Mathematics, Chinese Academy of Sciences,
Wuhan 430071, China}
\affiliation{University of the Chinese Academy of Sciences, Beijing 100049, China}
\author{Jian-Qi Zhang}
\email{changjianqi@gmail.com} \affiliation{State Key Laboratory of
Magnetic Resonance and Atomic and Molecular Physics, Wuhan Institute
of Physics and Mathematics, Chinese Academy of Sciences, Wuhan
430071, China}
\author{Shuo Zhang}
\affiliation{Zhengzhou Information Science and Technology Institute, Zhengzhou, 450004,
China}
\author{Mang Feng}
\email{mangfeng@wipm.ac.cn}
\affiliation{State Key Laboratory of
Magnetic Resonance and Atomic and Molecular Physics, Wuhan Institute
of Physics and Mathematics, Chinese Academy of Sciences, Wuhan
430071, China}

\begin{abstract}
Cooling vibrational degrees of freedom down to ground states is essential to
observation of quantum properties of systems with mechanical vibration. We
propose two cooling schemes employing four internal levels of the systems,
which achieve the ground-state cooling in an efficient fashion by completely
deleting the carrier and first-order blue-sideband transitions. The schemes,
based on the quantum interference and Stark-shift gates, are robust to
fluctuation of laser intensity and frequency. The feasibility of the schemes
is justified using current laboratory technology. In practice, our proposal readily
applies to an nano-diamond nitrogen-vacancy center levitated in an optic trap
or attached to a cantilever.
\end{abstract}

\maketitle



\noindent\textit{Introduction.}- To observe quantum characteristics of some
systems with spin-vibration coupling, e.g., trapped ions or nanomechanical
cantilevers, we have to eliminate the thermal phonons intrinsically owned by
the systems. As such, cooling a system down to ground states of the
vibrational degrees of freedom is usually a prerequisite of executing
quantum tasks, such as quantum computing \cite{rmp-75-281} and
ultra-precision measurements \cite%
{nature-430-329,pnas-106-1313,prl-108-120801}. Despite great
success so far in cooling of trapped ions or atoms, cooling solid-state
systems down to vibrational ground states is still tough experimentally.

Heating during the cooling process originates from two aspects. One
is from the environment. This heating in solid-state systems can be
suppressed by reducing the surface area of the system (or say,
enhancing the quality factor Q), decreasing the work temperature of
the surrounding environment \cite{nature-459-960,nature-464-697}
and/or dynamically controlling the dissipation
\cite{prl-110-153606}. The other is from some unexpected processes
in the cooling, such as the carrier and the blue-sideband
transitions. Suppressing those undesired transitions is an effective
way to the cooling acceleration \cite{prl-85-4458} . As such,
modified cooling schemes using quantum interference, such as the
electromagnetic induced transparency (EIT), can largely suppress the
blue-sideband transitions \cite
{nano-cooling-exp,nature-443-193,prl-85-4458} and work more
efficiently than the original idea of sideband cooling \cite
{prl-99-093901,prl-99-093902,nano-cooling-theory,rmp-86-1391}. In addition, a
proposal involving Stark-shift gate \cite{njp-9-279} can suppress
both carrier and blue-sideband transitions by steering the system to
red-sideband transitions.

In the present letter, we propose two efficient cooling schemes by
suppressing the unexpected processes as mentioned above, through employing
four internal levels of the systems. As clarified below, the schemes can
readily apply to vibrational systems involving nano-diamond nitrogen-vacancy
(nNV) centers. Due to large mass and special characteristic, the nNV center
system cannot be cooled down by simply merging previous cooling ideas, such
as the scheme with EIT plus Stark-shift-gate for cooling trapped ions \cite%
{prl-104-043003}. Magnetic field gradient, in addition to laser
irradiation, is required to couple the internal to the vibrational
degrees of freedom of the solid-state system. Besides, due to
existence of an additional decay to a metastable level in the nNV
center \cite{nature-466-730}, a pumping process in addition to usual
cooling operations is demanded \cite{oe-21-029695}. Even with all
these considerations, however, a three-level structure employed in a
nNV center is proven to be not qualified for a perfect cooling since
the first-order blue sideband transition cannot be fully eliminated
\cite {oe-21-029695,SR-5-14977,prb-79-041302}.

In contrast, the two schemes proposed in the present letter employ four
internal levels of the nNV center, which can well accomplish the cooling by
completely eliminating the carrier and the first-order blue sideband
transitions. One of our schemes with a modified $\Lambda $-type
configuration is based on a dynamical EIT \cite{prl-85-4458,oe-21-029695}
plus a Stark-shift gate \cite{njp-9-279,SR-5-14977}, which is called briefly
as the asymmetric cooling method. The other, called shortly as the symmetric
cooling method, combines the $\Lambda $-type with $V$-type configurations,
which yield double Stark-shift gates. As we know, most of the cooling
schemes proposed so far are based on $\Lambda $-type three-level systems
\cite{prl-85-4458,prl-103-227203, oe-21-029695,rmp-75-281}, rather than the $%
V$-type three-level configuration \cite{prb-79-041302} , due to the fact
that the latter with two upper levels is more susceptible to dissipation.
However, as shown below, a better cooling could be achieved if we
elaborately combine the $\Lambda $-type and $V$-type configurations and have
them interfered with each other. The interference enhanced by the
Stark-shift gates yields a dark state and a Stark-shift-gate point, which
help for an efficient cooling.

\begin{figure}[htbp]
\centering
\includegraphics[width=8cm,height=6.0cm]{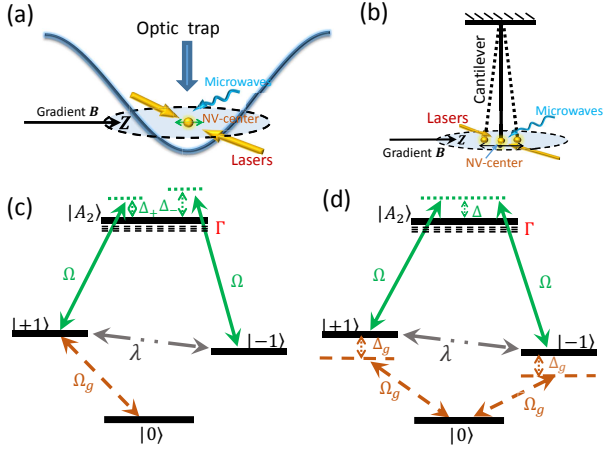}
\caption{(Color online) (a, b) Schematic illustrations of a single nNV
center levitated in an optic trap and attached at the head of a cantilever.
A magnetic field gradient is required to couple the internal state of the nNV
center to its own vibration in (a) or to the cantilever's vibration in (b). 
(c) The level configuration employed in the asymmetric cooling
method. The two levels $|\pm 1\rangle$ are coupled to the excited level $%
|A_2\rangle$ by a pair of laser beams with equal Rabi frequencies $\Omega$
but different detunings $\Delta_{\pm}$. A microwave resonantly couples the
lowest state $|0\rangle$ to $|+1\rangle$ with a Rabi frequency $\Omega_g$.
(d) The level configuration employed in the symmetric cooling method. The
two levels $|\pm 1\rangle$ are coupled to $|A_2\rangle$ by a pair
of laser beams with Rabi frequency $\Omega$ and detuning $\Delta$, and to $%
|0\rangle$ by a pair of microwaves with Rabi frequency $\Omega_g$ and
detuning $\Delta_g$. }
\label{Fig1}
\end{figure}

\noindent\textit{The systems.} - We exemplify two systems to clarify our
schemes. One is an nNV center levitated in an optic trap \cite%
{nphtonics-9-653} (see Fig. \ref{Fig1}(a)), which is promising for detecting
quantized gravity \cite{pra-90-033834}, preparing large distance
superpositions \cite{prl-107-020405,pra-88-033614} and building matter-wave
interferometers \cite{prl-111-180403}. The other is the nNV center attached
to a cantilever \cite{nphy-7-879} (see Fig. \ref{Fig1}(b)), which has
potential applications in observation of phononic Mollow triplet \cite%
{natcomm-6-8603}, ultra-sensitive measurements \cite{prl-108-120801},
quantum information processing \cite{prl-105-220501,natphys-6-602} and
biological sensing \cite{nnano-3-501}. To achieve the objectives, the
vibrational degrees of freedom in those systems are required to be cooled
down to ground states which are very challenging with current laboratory
technologies. We show below that our schemes can accomplish the cooling of
the above systems in an efficient and robust way.

The nNV center owns multiple levels and a small surface area, which ensure
it to be operated more flexibly and with lower heating from the environment
than other solid-state candidates. Besides, the internal electronic states
of the nNV center and the corresponding vibrational states can be coupled
strongly under a modest magnetic field gradient ($10^{5}$ T/m) \cite%
{science-335-1603}. Provided that the trapping frequency of the nNV center
(or the vibrational frequency of the cantilever) is $\omega_{k}$ and a
magnetic field gradient along the nNV center axis couples the nNV electron
spin ($S=1$) to its vibration, such a system can be described in units of $%
\hbar =1$ as $H_{0}=\omega_{k}a^{\dagger}a +\delta_B(|+1\rangle
\langle+1|-|-1\rangle \langle -1|)+\lambda (a^{\dagger }+a)(|+1\rangle
\langle+1|-|-1\rangle \langle -1|)$, where $a$ ($a^{\dagger}$) is the
annihilation (creation) operator of the vibration, $\delta_B=g_{e}\mu_{B}B(0)
$ is the energy difference between $|\pm 1\rangle$ induced by the magnetic
field and the coupling strength is $\lambda =g_{e}\mu_{B}B^{\prime}(0)z_{0}$
\cite{prb-79-041302} with the zero-point fluctuation amplitude $z_{0}=\sqrt{%
\hbar /(2M\omega_{k})}$, the mass of the NV center $M$, the Lande factor $%
g_{e}$, the Bohr magneton $\mu_{B}$, and the magnetic field gradient $%
B^{\prime }(0)$ along the nNV center axis. The ground state $|m_{s}\rangle$
with $m_{s}=0,\pm 1$ represents the Zeeman sublevels of the spin $S=1$. Due
to intrinsic spin-spin coupling properties, there exists a zero-field
splitting of $2\pi \times 2.88$ GHz between $|\pm 1\rangle$ and $|0\rangle$
\cite{science-314-281,prl-112-047601,prb-85-205203,njp-13-025025}.

\noindent\textit{The asymmetric cooling method.} - The nNV center is driven
by external light fields as in Fig. \ref{Fig1}(c). To carry out the cooling
scheme, we first employ a dark state, which, under the condition of
$\Delta_-=\Delta_{+}+\Omega_{g}/2$, is $|d\rangle =(|+1\rangle +|0\rangle-|-1\rangle)/\sqrt{3}$.
Correspondingly, we define the states $|b\rangle =(|+1\rangle
+|0\rangle+2|-1\rangle)/\sqrt{6}$ and $|Y\rangle=(|+1\rangle -|0\rangle)/
\sqrt{2}$. The steady state, obtained from the first-order expansion of the
parameter $\eta=\lambda/\omega_k$ \cite{prl-104-043003} can be written as
\begin{equation}
|\varphi\rangle=|d\rangle|0\rangle_{n}-\frac{\eta}{\sqrt{2}}(|b\rangle-\sqrt{%
3}|Y\rangle)|1\rangle_{n},  \label{Eq1}
\end{equation}
where $|j\rangle_{n}$ ($j=0,1$) denotes the phonon state. Normalization
factor is omitted in Eq. (\ref{Eq1}) for simplicity, but considered in
calculations below. The effective Hamiltonian contributing to the cooling is
$H^{a} = H_f+H_{\mathrm{EIT}}+H_{\mathrm{Stark}}$ with
\begin{eqnarray}
H_f &=&\omega_{k} a^{\dagger }a -\Delta_{-} |A_2\rangle \langle A_2|,
~~~~~~~~~~~~~~~~~~~~~~~~~~  \notag \\
H_{\mathrm{EIT}} &=& \frac{\sqrt{6}\Omega}{4}\sigma_x^{A_{2},b}+\frac{\sqrt{2%
}}{2}\lambda(a+a^{\dagger})\sigma_{x}^{b,d}, \\
H_{\mathrm{Stark}} &=&-\Omega_g |Y\rangle \langle Y|+\frac{\sqrt{2}\Omega}{4}%
\sigma_x^{A_{2},Y} + \frac{\sqrt{6}}{6}\lambda(a+a^{\dagger})%
\sigma_{x}^{d,Y},  \notag  \label{Eq2}
\end{eqnarray}
where $\sigma_{x}^{j,k}=|j\rangle \langle k|+|k\rangle \langle j|$ with $%
j,k=A_{2},Y,d,b$. $H_f$ denotes the free energy term, and $H_{\mathrm{EIT}%
}$ and $H_{\mathrm{Stark}}$ are terms employed in the EIT cooling and the
Stark-shift-gate cooling, respectively. So it is clear that the asymmetric
cooling method works with a combination of the EIT cooling and the
Stark-shift-gate cooling. For the steady state $|\varphi\rangle$, we have $%
H^a|\varphi\rangle=\sqrt{\frac{3}{2}} \eta(\frac{4\omega_k}{3}%
-\Omega_g)|Y\rangle|1\rangle_n$, which remains invariant under the condition
\begin{equation}
3\Omega_g=4\omega_k.  \label{Eq3}
\end{equation}
This condition, also called Stark-shift-gate point, implies that all the
terms in $H^{a}$ act on the steady state by a destructive interference and
thus the steady state does not suffer from any loss due to spontaneous
emission.

\noindent\textit{The symmetric cooling method.} - With respect to
the asymmetric cooling method, we apply an additional microwave to
couple the ground state $|0\rangle$ to the ground state
$|-1\rangle$, as shown in Fig. \ref{Fig1}(d). The level
configuration in this case is actually a combination of the
$\Lambda$-type and $V$-type structures, which looks graphically
symmetric. In this case, it is easily proven that $
|D\rangle=(|+1\rangle -|-1\rangle)/\sqrt{2}$ is the dark state under
the double resonant conditions. The Hamiltonian here is written as
$H^s=H_0^s+H_{\mathrm{Stark}}+H^{\prime}_{\mathrm{Stark}}$, with
\begin{eqnarray}
H_0^s&=&\omega_k a^{\dagger}a+\lambda (a+a^{\dagger })(|B\rangle \langle
D|+|D\rangle\langle B|),  \notag \\
H_{\mathrm{Stark}}&=&-\Delta (|A_2\rangle\langle A_2|-|B\rangle\langle B|)+
\frac{\sqrt{2}\Omega}{2}\sigma_{x}^{A_2,B}, ~~~~~~~  \notag \\
H^{\prime}_{\mathrm{Stark}}&=&(\Delta _{g}|0\rangle \langle 0|-\Delta |B
\rangle\langle B|)+\frac{\sqrt{2}\Omega_g }{2}\sigma_x^{0,B},  \label{Eq4}
\end{eqnarray}
where the state $|B\rangle =(|+1\rangle +|-1\rangle)/\sqrt{2}$. Eq. (\ref%
{Eq4}) indicates that the symmetric cooling method works with two
Stark-shift-gate cooling processes collaboratively. The first-order
expansion of $\eta$ for the steady state $|\phi\rangle$ (omitting the
normalization factor) \cite{prl-104-043003} is $|\phi\rangle=|D\rangle|0%
\rangle_n-\frac{\sqrt{2}\eta\omega_k} {\Omega_g}|0\rangle|1\rangle_n $.
Similar to the asymmetric cooling method, we can have a simplified
expression from $H^s$ applied on $|\phi\rangle$, $H^s|\phi\rangle=-\frac{%
\sqrt{2}\eta\omega_k}{\Omega_g}(\omega_k+\Delta_g)|0\rangle|1\rangle_n $
which is invariant once the detuning $\Delta_g$ meets the condition of
Stark-shift-gate point,
\begin{equation}
\Delta_g=-\omega_k.  \label{Eq5}
\end{equation}

\begin{figure}[htbp]
\centering
\includegraphics[width=9.2cm,height=4cm]{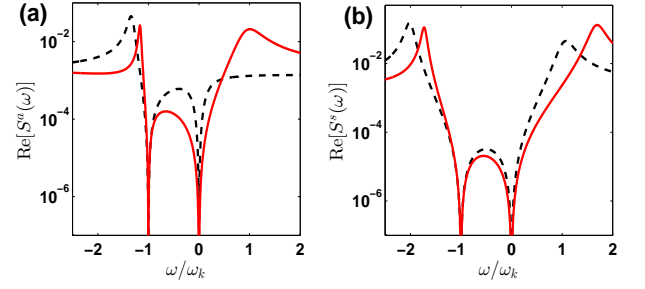}
\caption{(Color online) (a) and (b) Fluctuation spectra for the asymmetric
and symmetric cooling schemes, respectively. The dashed (solid) curves are
plotted with $\Omega/\protect\omega_k=2.0 (8.0)$ and $\eta=0.1$.}
\label{Fig2}
\end{figure}

\noindent\textit{The cooling effects.} - We first solve the cooling
coefficients by the fluctuation spectrum, which is given by \cite{spra-46-2668}
\begin{equation}
S(\omega )=\frac{1}{2M\omega }\int_{0}^{\infty }dte^{i\omega t}\langle
F(t)F(0)\rangle_{\mathrm{ss}},  \label{Eq6}
\end{equation}
where $\langle\cdots\rangle _{\mathrm{ss}}$ implies the expectation
value over the steady state. In the Schr{\"o}dinger representation, the operator $
F=-(d/dz)H_{\mathrm{int}}|_{z=0}$ with $z=z_0 (a+a^{\dagger})$
and $H_{\mathrm{int}}= \lambda (a^{\dagger }+a)(|+1\rangle
\langle+1|-|-1\rangle \langle -1|)$. For the steady state $ |\varphi\rangle$
in the asymmetric cooling scheme, we find two forces, $F_{
\mathrm{EIT}}=-(\eta\omega_k/\sqrt{2}z_0)\sigma_{x}^{d,b}$ and
$F_{\mathrm{Stark}}=-(\eta\omega_k/\sqrt{6}z_0)\sigma_{x}^{D,d}$,
contributing to the cooling. These two forces split the fluctuation
spectrum $S^a(\omega)$ into three parts, that is,
$S_{\mathrm{EIT}}(\omega)$, $S_{\mathrm{Stark}}(\omega)$ and their
interaction term $S_{\mathrm{int}}(\omega)$ \cite{supplementary
information}. The heating coefficient can be obtained by the
fluctuation spectrum as
$A_{+}=2\text{Re}[S^a(-\omega_k)]=\Gamma\eta^2\omega_k^2\Omega^2
(3\Omega_g-4\omega_k)^2/12|M(-\omega_k)|^2$ with
$M(\omega)=(-3\Omega^2/4+2 \Delta
\omega+2\omega^2)(\Omega_g+\omega)-\omega\Omega^2/4+i\Gamma\omega(\omega+
\Omega_g)$ \cite{supplementary information}. Under the condition in
Eq. (\ref {Eq3}), we have $A_{+}=0$ and zero values in fluctuation spectra
at the points $\omega/\omega_k=-1, 0$ (see Fig.
\ref{Fig2}(a)), which means that quantum interference induced by the
EIT and the Stark-shift gate fully eliminates heating from the
first-order blue sideband and carrier transitions. At the detuning
$\Delta=(3\Omega^2/7\omega_k)-\omega_k$, the cooling coefficient
reaches the peak value
\begin{equation}
A^{a}_{-}=2\text{Re}[S^a(\omega_k)]=\frac{48\lambda ^{2}\Omega ^{2}}{%
49\Gamma \omega _{k}^{2}},  \label{Eq7}
\end{equation}
which is similar to the square-law style of the cooling laser strength in
the EIT cooling scheme \cite{oe-21-029695}.

Similar to the asymmetric cooling scheme, the fluctuation spectra in
symmetric counterpart are obtained and plotted in Fig. \ref{Fig2}(b). The
heating coefficient in this scheme reaches zeros if Eq. (\ref{Eq5}) is
satisfied. When we choose $\Omega_{g}=2\omega_{k}$, a maximal cooling
coefficient is obtained as
\begin{equation}
A_{-}^{s}=2\text{Re}[S^s(\omega_k)]=\frac{2\Gamma \lambda ^{2}}{\Omega ^{2}},
\label{Eq8}
\end{equation}
which obeys the inverse square-law as in \cite{njp-9-279,SR-5-14977}.
Evidently, a better cooling by this scheme favors weaker lasers.

\begin{figure}[tbph]
\centering
\includegraphics[width=3.5cm,height=7cm]{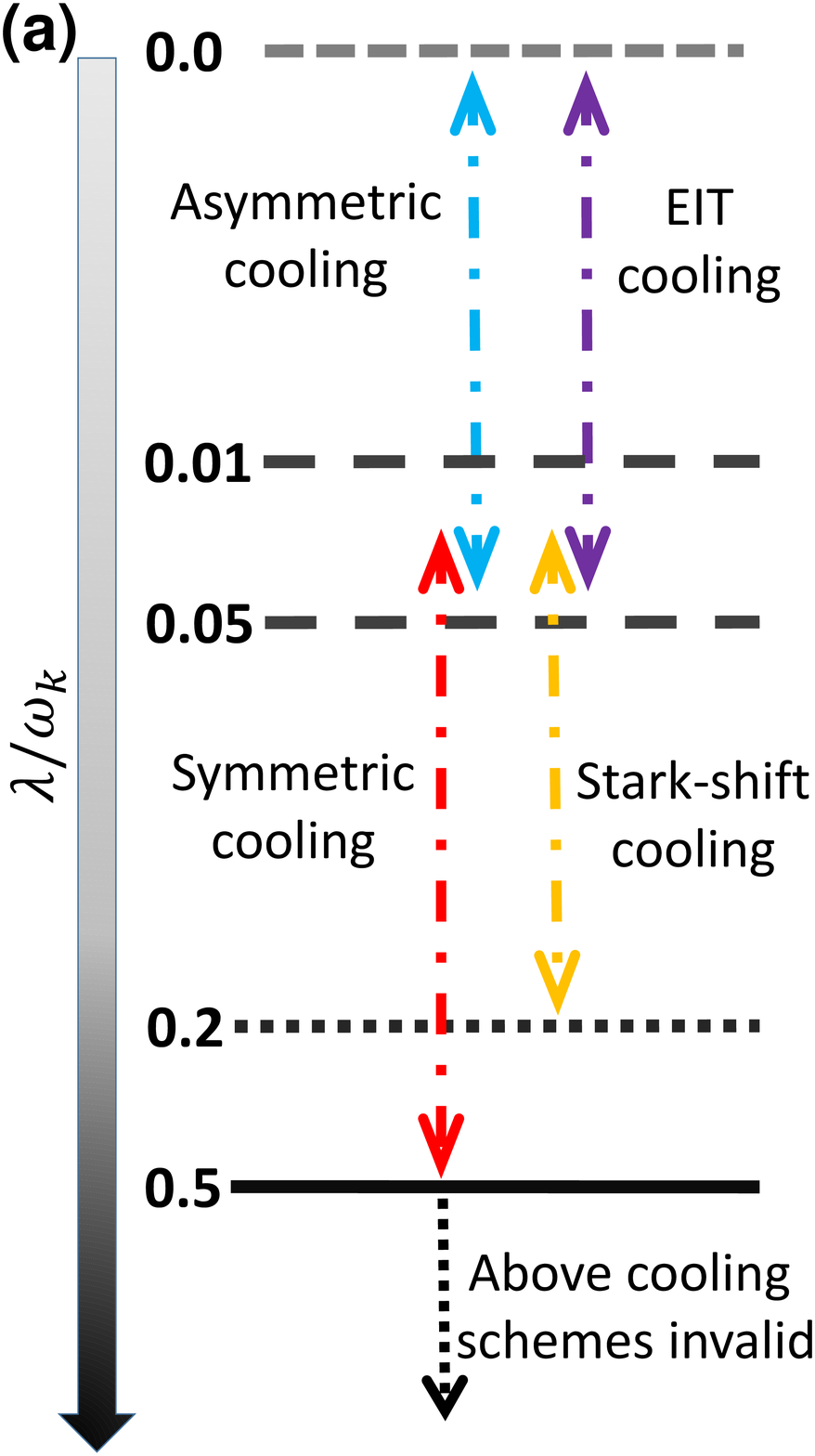} \quad 
\includegraphics[width=4.6cm,height=7cm]{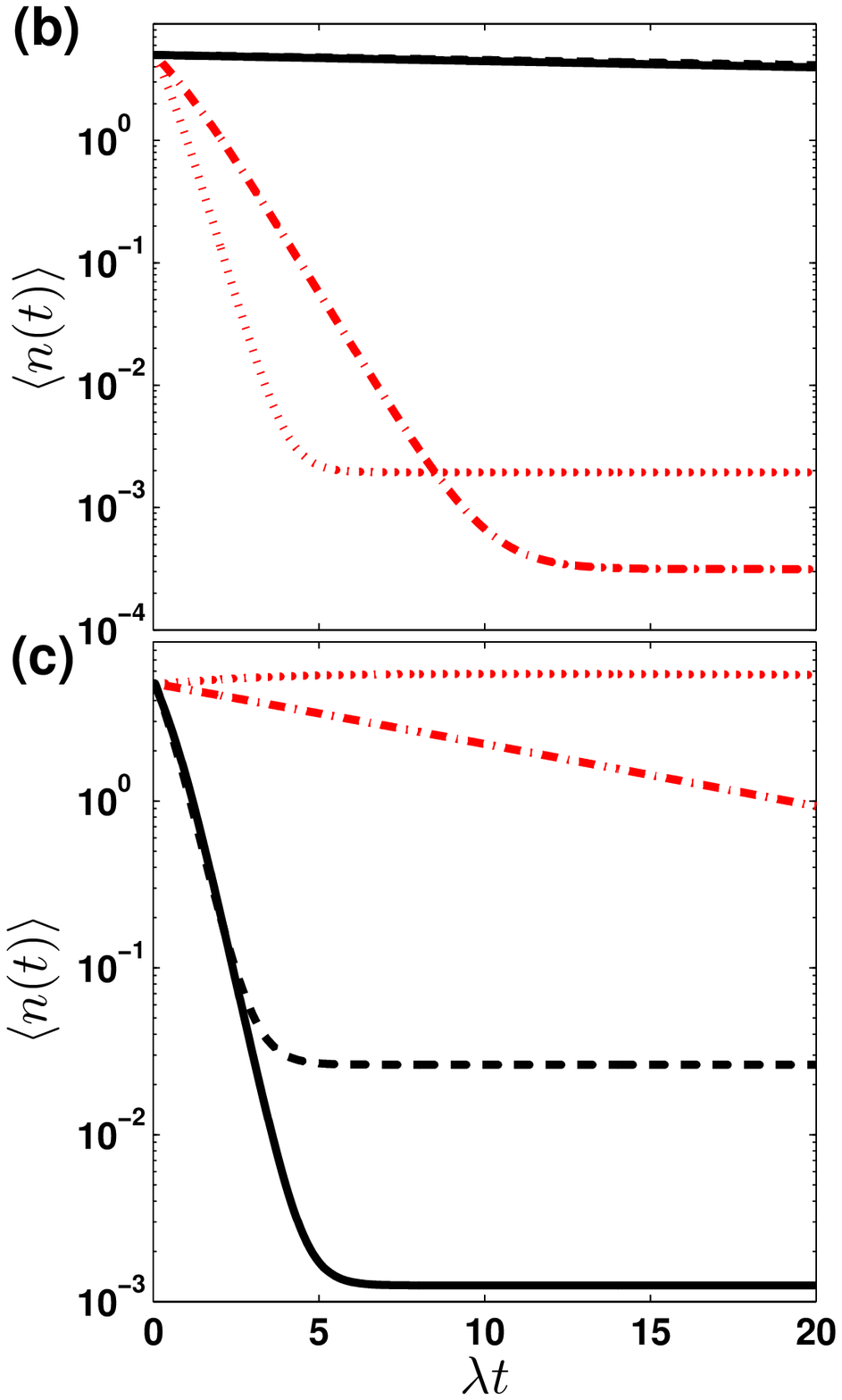}
\caption{(Color online) (a) Sketch for working regions of typical cooling
schemes in the change of $\protect\lambda/\protect\omega_{k}$ under the
assumption of appropriate parameter values. (b) and (c) Dynamics of $\langle
n(t)\rangle$ in execution of these schemes in different regions, where (b) $%
\eta=\protect\lambda/\protect\omega_k=1/80$ with $\Omega/\protect\omega_k=5.0$
and (c) $\eta=1/20$ with $\Omega/\protect\omega%
_k=1.5$. In each panel, the dashed-dotted, dotted, dashed and solid curves
represent the asymmetric cooling, EIT \protect\cite{oe-21-029695},
Stark-shift-gate \protect\cite{njp-9-279} and the symmetric cooling schemes,
respectively. Here we suppose zero-temperature environment and the decay
rate of the internal levels to be $\Gamma=150\protect\lambda$. Other
parameters are chosen to satisfy the destructive interference and the
resonance condition. Values would be changed for a finite-temperature
environment, but physical essence as expressed here remains. }
\label{Fig3}
\end{figure}

To demonstrate the cooling effects more specifically, we simulate the
dynamics of the system under a zero-temperature environment by the Lindblad
master equation $d\rho/dt=\mathcal{L}_0\rho$ where
\begin{equation}
\mathcal{L}_0=-i[H,\rho]+\underset{i=\pm 1,0}{\sum }\frac{\Gamma _{i}}{2}%
\mathcal{D}[|i\rangle \langle A_{2}|,\rho],  \label{Eq9}
\end{equation}
with $\mathcal{D}[o,\rho ^{j}]=2o\rho ^{j}o^{\dagger
}-o^{\dagger}o\rho^{j}-\rho ^{j}o^{\dagger}o$. We find working regions of
different cooling schemes, as sketched in Fig. \ref{Fig3}a. Although the
asymmetric cooling scheme favors a weaker coupling (i.e., smaller values of $%
\lambda$), stronger lasers (i.e., larger values of $\Omega$) are necessary
in that case. As plotted in Fig. \ref{Fig3}b, the asymmetric scheme can
achieve the lowest final phonon number among the four typical schemes when $%
\eta<$0.05. In contrast, if we have a stronger coupling (e.g.,
with bigger magnetic field gradient), we may achieve the cooling by the
symmetric scheme using weaker lasers (see Fig. \ref{Fig3}c). An evident
reason for this difference between the two schemes is due to different forms
of $A_{-}$ as in Eqs. (\ref{Eq7}) and (\ref{Eq8}) as well as the relation of
the cooling coefficient with the final phonon number, i.e.,
$\langle n\rangle_{\mathrm{ss}}\sim 1/A_{-}$ in the case of $A_{+}=0$.
Deeper physics for this difference is reflected in different
properties between the EIT cooling and the Stark-shift-gate cooling. In the asymmetric
scheme, the $\Lambda$-type structure plus the Stark-shift gate constitutes an enhanced EIT
cooling that is steered to a better cooling rate under the condition of
$\Omega\sim \sqrt{\Gamma\omega_k/\eta}$. In contrast, the symmetric cooling is essentially
an enhanced Stark-shift cooling, in which the better cooling occurs if $\Omega\sim \sqrt{%
\eta\Gamma\omega_k}$. Since $\eta$ is always smaller than 0.5, stronger lasers are definitely
required in the asymmetric cooling scheme.
Due to the differences in physics, once we switch from the asymmetric
scheme to the symmetric one by applying an additional microwave, although we
may have a sudden transition from $H^{a}$ to $H^{s}$, cooling rate would
significantly drop down unless other conditions, such as the appropriate
laser strength, are also satisfied.

Under a realistic circumstance, the nNV center is influenced by a
finite-temperature environment, where the phonon occupations in the
vibrational degrees of freedom satisfy Bolzmann distribution as $%
N(\omega_{k})=[\exp (\omega_{k}/T)-1]^{-1}$ in units of $k_B=1$. As such, an
additional Lindblad operator $\mathcal{L}_1$ should be involved in the
Lindblad master equation with
\begin{equation*}
\mathcal{L}_1\rho=\frac{N(\omega _{k})\gamma _{k}}{2}\mathcal{D}[b^{\dagger
},\rho]+\frac{[N(\omega _{k})+1]\gamma _{k}}{2}\mathcal{D}[b,\rho].
\end{equation*}
Considering the vibrational decay rate $\gamma_{k}=\omega_{k}/Q$ with the
quality factor $Q$ \cite{prl-85-4458,prl-103-227203,
oe-21-029695,pra-67-033402}, the final average number of phonons $\langle
n\rangle _{\mathrm{ss}}=[A_{+}+N(\omega_{k})\gamma_{k}]/(W+\gamma_{k})$ with
the cooling rate $W=A_--A_+$ can be obtained by solving the rate equation
\cite{prl-103-227203,HP-FP}. The term $N(\omega_{k})\gamma_{k}$ turns to be
most important in determining $\langle n\rangle _{\mathrm{ss}}$ when $A_+=0$%
. Ideally, the final average phonon number is $\langle n\rangle_{\mathrm{ss}%
}=A_+/W$, which vanishes if $A_+=0$.

\noindent\textit{Robustness.} - A working scheme for cooling should be
highly robust to the parameter fluctuation, which is essential to
experimental implementation. In our case, the deviations from the conditions
of Eqs. (3) and (5) yield second-order effects away from the ideal final
average phonon numbers, i.e., $\delta^a_{\langle n\rangle_{\mathrm{ss}%
}}\varpropto (\delta_{\Omega_g})^2$ and $\delta^s_{\langle n\rangle_{\mathrm{%
ss}}}\varpropto (\delta_{\Delta_g})^2$ \cite{supplementary information}. In
contrast, the EIT cooling and the Stark-shift-gate cooling are more
sensitive to such deviations, whose effect is reflected in the first-order
expansion. Further considering the inaccuracy of $\Omega$ in our schemes, we
find $\delta^a_{\langle n\rangle_{\mathrm{ss}}}\varpropto
\delta_{\Omega}(\delta_{\Omega_g})^2$ and $\delta^s_{\langle n\rangle_{%
\mathrm{ss}}}\varpropto\delta_{\Omega} (\delta_{\Delta_g})^2$ \cite%
{supplementary information}, i.e., only the third-order effect in the
cooling.

\noindent\textit{Experimental feasibility.} - For a nNV center levitated in
an optic trap \cite{pra-90-033834,prl-111-180403, pra-88-033614}, the
diameter of the nNV sphere is about 20 - 200 nm, the vibrational frequency
is $\omega_k/2\pi\approx 100 \sim$ 500 kHz, and the magnetic field gradient $%
G\approx 10^3\sim 10^5$ T/m and the coupling strength $\lambda/2\pi\le 100$
kHz are available. Moreover, the mechanical quality factor Q is relevant to
the pressure of the vacuum and can be as high as $3\times 10^{12}$ for a
pressure $P=10^{-10}$ Torr \cite{pnas-107-1005}. The decay from $|A_2\rangle$
is $\Gamma/2\pi=15$ MHz and the environment can be at room temperature \cite%
{pra-88-033614,pnas-107-1005} or cryogenic (such as 1 mK) \cite%
{pra-90-033834}. For a specific calculation, we choose $\omega_k/2\pi=500$
kHz and $\lambda/2\pi=50$ kHz, implying within the working region of the
symmetric cooling scheme. Provided that the environment is at room
temperature T$=300 $K, $Q=10^{10}$ and $\Omega/2\pi=1.5$ MHz, we may achieve
$\langle n\rangle_{ss}\approx 0.027$ after cooling for 100 $\mu$s, better
than the cooling by the Stark-shift-gate scheme (only reaching $\langle
n\rangle_{ss}\approx 0.082$).

With respect to the symmetric cooling scheme, the asymmetric scheme favors a
smaller $\eta$ and meanwhile saves a microwave irradiation.
For a nNV center attached at the end of a cantilever, since the vibrational
frequency of the cantilever is usually of the order of few MHz or larger,
the asymmetric scheme applies to such a case. Provided that the
environmental temperature T$=20$ mK, $\omega_k/2\pi= 8$ MHz, $Q=10^{6}$, $%
\lambda/2\pi=500$ kHz and $\Omega/2\pi=40$ MHz, we may cool the cantilever's
vibration by the asymmetric scheme down to $\langle n\rangle_{ss}\approx
0.013$ after cooling for 90 $\mu$s.

\noindent\textit{Discussion.} - Two special characteristics of the NV
center, as the difference from the atoms', should be mentioned. First, there
is a leaking channel out of the four-level configuration employed in Fig. %
\ref{Fig1}(c, d), i.e., from $|A_{2}\rangle$ to $|^{1}A\rangle$, and finally
down to $|0\rangle$. For the three-level structure involving $|\pm 1\rangle$
and $|A_{2}\rangle$, a pumping from $|0\rangle$ back to $|\pm 1\rangle$ is
required in order to accomplish the desired cooling \cite{oe-21-029695}. 
In contrast, there is no need to additionally consider such a pumping in 
the present schemes since $|0\rangle$ is
involved in the four levels. Second, the detrimental influence from nuclear
spin bath of $^{14}N$ and $^{13}C$ should be seriously considered in the nNV
center. This unexpected effect is neglected in above treatment, but actually
leads to slight deviation from the dark-state condition and to inefficiency
in removing the first-blue sideband transition. Simple estimate involving
the nuclear spin noise can be found in the Supplementary Material \cite%
{supplementary information}. It is evident that this noise is beyond the
scope of our schemes and should be suppressed by other approaches \cite%
{prl-102-057403,nl-12-2083}.

\noindent\textit{Summary.}- Our proposed efficient cooling schemes, based on
quantum interference enhanced by the Stark-shift gates, can achieve a very
low final average phonon number in some hot-topic systems with nNV centers.
To the best of our knowledge, the schemes are the first proposal for cooling
solid-state systems by completely eliminating heating effects from
carrier transitions and the first-order blue sideband transitions. The two
schemes favor slightly different conditions, and if employed appropriately,
can be very general to achieve cooling for systems with four internal levels
and a wide range of vibrational frequencies. Besides the two systems
exemplified above, our cooling schemes can readily apply to other
spin-vibration coupling systems involving four internal levels.

This work is supported by National Fundamental Research Program of China
under Grants No. 2012CB922102 and No. 2013CB921803, and by National Natural
Science Foundation of China under Grants No. 11274352 and No. 11304366.


\begin{thebibliography}{99}
\bibitem{rmp-75-281} D. Leibfried, R. Blatt, C. Monroe, and D. Wineland,
Rev. Mod. Phys. \textbf{75}, 281 (2003).

\bibitem{nature-430-329} D. Rugar, R. Budakian, H. J. Mamin, and B. W. Chui,
Nature \textbf{430}, 329 (2004).

\bibitem{pnas-106-1313} C. L. Degen, M. Poggio, H. J. Mamin, C. T. Rettner,
and D. Rugar, Proc. Natl. Acad. Sci. USA \textbf{106}, 1313 (2009).

\bibitem{prl-108-120801} S. Forstner, S. Prams, J. Knittel, E. D. van
Ooijen, J. D. Swaim, G. I. Harris, A. Szorkovszky, W. P. Bowen, and H. R.
Dunlop, Phys. Rev. Lett. \textbf{108}, 120801 (2012).

\bibitem{nature-459-960} M. D. LaHaye, J. Suh, P. M. Echternach, K. C.
Schwab, and M. L. Roukes, Nature (London) \textbf{459}, 960 (2009).

\bibitem{nature-464-697} A. D. O'Connell, M. Hofheinz, M. Ansmann, R.
C. Bialczak, M. Lenander, E. Lucero, M. Neeley, D. Sank, H. Wang, M.
Weides, J. Wenner, John M. Martinis and A. N. Cleland, Nature (London) \textbf{464}, 697
(2010).

\bibitem{prl-110-153606} Y.-C. Liu, Y.-F. Xiao, X.-S. Luan, and C. W. Wong,
Phys. Rev. Lett. \textbf{110}, 152606 (2013).

\bibitem{nano-cooling-exp} J. D. Teufel, T. Donner, D. Li, J. W. Harlow,
M. S. Allman, K. Cicak, A. J. Sirois, J. D. Whittaker, K. W. Lehnert, and R.
W. Simmonds, Nature (London) \textbf{475}, 359 (2011).

\bibitem{nature-443-193} A. Naik, O. Buu, M. D. LaHaye, A. D. Armour, A. A.
Clerk, M. P. Blencowe, and K. C. Schwab, Nature (London) \textbf{443}, 193
(2006).

\bibitem{prl-85-4458} G. Morigi, J. Eschner, and C. H. Keitel, Phys. Rev.
Lett. \textbf{85}, 4458 (2000).

\bibitem{prl-99-093901} I. Wilson-Rae, N. Nooshi, W. Zwerger, and T. J.
Kippenberg, Phys. Rev. Lett. \textbf{99}, 093901 (2007).

\bibitem{prl-99-093902} F. Marquardt, J. P. Chen, A. A. Clerk, and S. M. Girvin,
Phys. Rev. Lett. \textbf{99}, 093902 (2007)


\bibitem{nano-cooling-theory} I. Wilson-Rae, P. Zoller, and A. Imamo{\v g}%
lu, Phys. Rev. Lett. \textbf{92}, 075507 (2004).

\bibitem{rmp-86-1391} M. Aspelmeyer, T. J. Kippenberg, and F. Marquardt,
Rev. Mod. Phys. \textbf{86}, 1391 (2014).

\bibitem{njp-9-279} A. Retzker and M. B. Plenio, New J. Phys. \textbf{9},
279 (2009).

\bibitem{prl-104-043003} J. Cerrillo, A. Retzker, and M. B. Plenio, Phys.
Rev. Lett. \textbf{104}, 043003 (2010).

\bibitem{nature-466-730} E. Togan, Y. Chu, A. S. Trifonov, L. Jiang, J.
Maze, L. Childress, M. V. G. Dutt, A. S. S$\o$rensen, P. R. Hemmer, Nature
(London) \textbf{466}, 730 (2010).

\bibitem{oe-21-029695} J.-Q. Zhang, S. Zhang, J.-H. Zou, L. Chen, W. Yang,
Y. Li, and M. Feng, Opt. Exp. \textbf{21}, 029695 (2013).

\bibitem{SR-5-14977} L.L. Yan, J. Q. Zhang, S. Zhang, and M. Feng, Sci. Rep.
\textbf{5}, 14977 (2015).

\bibitem{prb-79-041302} P. Rabl, P. Cappellaro, M. V. G. Dutt, L. Jiang, J.
R. Maze, and M. D. Lukin, Phys. Rev. B \textbf{79}, 041302 (2009).

\bibitem{prl-103-227203} K.-Y. Xia and J{\"o}rg Evers, Phys. Rev. Lett.
\textbf{103}, 227203 (2009).

\bibitem{nphtonics-9-653} L. P. Neukirch, E. von Haartman, J. M. Rosenholm,
and A. N. Vamivakas, Nat. Photo. \textbf{9}, 653 (2015).

\bibitem{pra-90-033834} A. Albrecht, A. Retzker, and M. B. Plenio, Phys.
Rev. A \textbf{90}, 033834 (2014).

\bibitem{prl-107-020405} O. Romero-Isart, A. C. Pflanzer, F. Blaser, R.
Kaltenbaek, N. Kiesel, M. Aspelmeyer, and J. I. Cirac, Phys. Rev. Lett.
\textbf{107}, 020405 (2011).

\bibitem{pra-88-033614} Z. Q. Yin, T. C. Li, X. Zhang, and L. M. Duan, Phys.
Rev. A \textbf{88}, 033614 (2013).

\bibitem{prl-111-180403} M. Scala, M. S. Kim, G. W. Morley, P. F. Barker,
and S. Bose, Phys. Rev. Lett. \textbf{111}, 180403 (2013).

\bibitem{nphy-7-879} O. Arcizet, V. Jacques, A. Siria, P. Poncharal, P.
Vincent, and S. Seidelin, Nat. Phys. \textbf{7}, 879 (2011).

\bibitem{natcomm-6-8603} B. Pigeau, S. Rohr, L. M. de L{\' e}pinay, A.
Gloppe, V. Jacques, and O. Arcizet, Nat. Commun. \textbf{6}, 8603 (2015).

\bibitem{prl-105-220501} K. Stannigel, P. Rabl, A. S. Sorensen, P. Zoller,
and M. D. Lukin, Phys. Rev. Lett. \textbf{105}, 220501 (2010).

\bibitem{natphys-6-602} P. Rabl, S. J. Kolkowitz, F. H. L. Koppens, J. G. E.
Harris, P. Zoller, and M. D. Lukin, Nat. Phys. \textbf{6}, 602 (2010).

\bibitem{nnano-3-501} L. Tetard, A. Passian, K. T. Venmar, R. M. Lynch, B.
H. Voy, G. Shekhawat, V. P. Dravid, and T. Thundat, Nat. Nanotechnol.
\textbf{3}, 501 (2008).

\bibitem{science-335-1603} S. Kolkowitz, A. C. B. Jayich, Q. P.
Unterreithmeier, S. D. Bennett, P. Rabl, J. G. E. Harris, and M. D. Lukin,
Science \textbf{335}, 1603 (2012).

\bibitem{science-314-281} L. Childress, M. V. G. Dutt, J. M. Taylor, A. S.
Zibrov, F. Jelezko, J. Wrachtrup, P. R. Hemmer, and M. D. Lukin, Science,
\textbf{314}, 281 (2006).

\bibitem{prl-112-047601} M. W. Doherty, V. V. Struzhkin, D. A. Simpson, L.
P. McGuinness, Y.-F. Meng, A. Stacey, T. J. Karle, R. J. Hemley, N. B.
Manson, L. C. L. Hollenberg, and S. Prawer, Phys. Rev. Lett. \textbf{112},
047601 (2014).

\bibitem{prb-85-205203} M. W. Doherty, F. Dolde, H. Fedder, F. Jelezko, J.
Wrachtrup, N. B. Manson, and L. C. L. Hollenberg, Phys. Rev. B \textbf{85},
205203 (2012).

\bibitem{njp-13-025025} J. R. Maze, A. Gali, E. Togan, Y. Chu, A. Trifonov,
E. Kaxiras, and M. D. Lukin, New J. Phys. \textbf{13}, 025025 (2011).

\bibitem{spra-46-2668} J. I. Cirac, R. Blatt, and P. Zoller, Phys. Rev. A
\textbf{46}, 2668 (1992).

\bibitem{supplementary information} See details in Supplementary Material.

\bibitem{pra-67-033402} G. Morigi, Phys. Rev. A \textbf{67}, 033402 (2003).

\bibitem{HP-FP} H.-P Breuer and F. Petruccione, \textit{The theory of open
quantum system}, Oxford, 2002.

\bibitem{pnas-107-1005} D. E. Chang, C. A. Regal, S. B. Papp, D. J. Wilson,
J. Ye, O. Painter, H. J. Kimble, and P. Zoller, Proc. Natl. Acad. Sci. USA
\textbf{107}, 1005 (2010).

\bibitem{prl-102-057403} V. Jacques, P. Neumann, J. Beck, M. Markham, D.
Twitchen, J. Meijer, F. Kaiser, G. Balasubramanian, F. Jelezko, and J.
Wrachtrup, Phys. Rev. Lett. \textbf{102}, 057403 (2009).

\bibitem{nl-12-2083} T. Ishikawa, K.-M. C. Fu, C. Santori, V. M. Acosta, R.
G. Beausoleil, H. Watanabe, S. Shikata, and K. M. Itoh, Nano. Lett. \textbf{%
12}, 2083 (2012).
\end{thebibliography}
\end{document}